# Designing Homogeneous Ti-Nb-Fe-Sn β Titanium Alloys by PBF-LB: A Pre-Alloyed Powder Blend Strategy

João Felipe Queiroz Rodrigues[a*], Kristína Bartha[b], Mariano Casas-Luna[b], Gilberto Vicente Prandi[a], Márcio Sangali[c], Kateřina Ficková[b], Jiří Kozlík[b], Michaela Šlapáková[b], Martin Koller[d], Adam Strnad[b], Josef Stráský[b], Miloš Janeček[b], Rubens Caram[a]

[a] Departamento de Engenharia de Manufatura e Materiais, Faculdade de Engenharia Mecânica, Universidade Estadual de Campinas (UNICAMP), 13083-970, Campinas, SP, Brazil

[b] Department of Physics of Materials, Faculty of Mathematics and Physics, Charles University, Ke Karlovu 5, 121 16, Prague, Czech Republic

[c] Federal Institute of Education, Science and Technology of São Paulo (IFSP), São João da Boa Vista, SP, 13871-298, Brazil

[d] Institute of Thermomechanics, Czech Academy of Sciences, Dolejškova 1402/5, 182 00, Prague, Czech Republic

## ABSTRACT

Metastable β titanium alloys are attractive for biomedical and structural applications owing to their low elastic modulus, high specific strength, and excellent corrosion resistance. Laser powder bed fusion (PBF-LB) enables complex-shape production and controlled compositional variation through powder blending. However, processing elemental Ti-Nb blends often results in chemical heterogeneity from incomplete dissolution of Nb-rich particles and non-equilibrium phase formation. To address this, low-modulus Ti-Nb-Fe-Sn alloys were produced by PBF-LB using Ti-42Nb, Ti-20Nb-15Fe, and Ti-20Nb-20Sn master-alloy powders blended with commercially pure Ti. Ti-23Nb-3Fe-4Sn, Ti-26Nb-2Fe-4Sn, Ti-29Nb-1Fe-4Sn, and Ti-32Nb-4Sn were fabricated using an uncommonly large 70 μm layer thickness with layer remelting, followed by heat treatment at 1000 °C for 2 h and water quenching. After heat treatment, all alloys exhibited low porosity, homogeneous chemical distribution, and single β-phase microstructures with predominantly equiaxed grains and weak crystallographic texture. Thermodynamic calculations indicated that solidification descriptors alone could not explain the non-monotonic grain-size evolution, which was attributed to inherited solidification structure, transient TiFe-like phase formation, Nb/Sn partitioning, and/or solute-drag-controlled β-grain growth. Hardness and yield strength decreased with decreasing Fe and increasing Nb contents, from 268 to 224 HV and 691 to 468 MPa, respectively. Young's modulus, determined by resonant ultrasound spectroscopy, ranged from 63 to 81 GPa. These results demonstrate that pre-alloyed master-alloy blends combined with remelting and heat treatment provide an effective route for producing chemically homogeneous Ti-Nb-Fe-Sn β alloys while revealing how small compositional changes govern grain-growth behavior and mechanical response.

## 1. Introduction

Titanium alloys are widely used in biomedical and structural applications due to their high specific strength, corrosion resistance, and biocompatibility. Among them, metastable β-type titanium alloys have attracted particular attention for load-bearing implants, as the retention of the body-centered cubic β phase generally provides a lower elastic modulus than conventional α+β alloys, such as Ti-6Al-4V, thereby reducing the risk of stress shielding and bone resorption around the implant [1,2]. Depending on composition and thermomechanical history, metastable β titanium alloys can also exhibit transformation-assisted deformation, superelasticity, or shape-memory response, expanding their potential beyond purely structural applications [1,2].

Ti-Nb-based alloys are particularly attractive because Nb is considered a biocompatible and an effective β-stabilizing element in titanium [1,2]. The addition of Fe offers a cost-efficient and production-efficient route to distribute β-phase stabilization across multiple solutes rather than relying exclusively on high Nb contents, which might result in complications with incomplete Nb dissolution [3]. Fe is a potent β stabilizer and solid-solution strengthener, and recent studies have demonstrated that Fe-containing titanium alloys processed by additive manufacturing can achieve attractive strength–ductility combinations when solute distribution is properly controlled [3,4]. Sn is also relevant in Ti-Nb-based alloys because it can contribute to solid-solution strengthening, and probably suppress or retard ω-phase formation, which is often associated with increased modulus and embrittlement in metastable β titanium alloys [1,5,6]. Thus, Ti-Nb-Fe-Sn alloys provide a compositional platform in which β stability, mechanical strength, and phase-transformation can be tuned simultaneously [3]. Finally, Ti-Nb-Fe-Sn alloying system allow for significant variations of mechanical properties, namely strength and elastic modulus, via minor differences in chemical composition.

Laser powder bed fusion (PBF-LB) provides a versatile tool for producing functionally graded materials [7], and layered structures of controlled heterostructures [8,9] based on variation of chemical composition. Such compositional variation is usually achieved by mixing elemental or pre-alloyed powders either before actual deposition in several hoppers or during deposition by manipulating the powders feeding rate. However, even when producing an intentionally gradient or layered product, the material must be locally (at the length-scale of initial powder particles) homogeneous to avoid formation of embrittling non-equilibrium phases or uncontrolled inclusions resulting in failure.

Successful PBF-LB production of homogeneous materials with various chemical compositions is of utmost importance and a clear challenge in the case of Ti-Nb-Fe-Sn alloys. Producing chemically homogeneous Ti-Nb alloys remains challenging when elemental powder blends are used [10,11]. The high melting point of Nb, the short lifetime of the melt pool, and the sluggish solid-state diffusion of Nb in Ti can lead to partially dissolved or unmelted Nb-rich particles, microsegregation, and local variations in phase stability [12–15]. These heterogeneities are especially critical in metastable β alloys because small local changes in composition may promote α'/α'' martensite, or ω formation, leading to spatial variations in hardness, elastic modulus, ductility, and damage tolerance [13,16].

A common strategy to improve dissolution of refractory particles during PBF-LB is to increase the local energy input, for example by using higher laser power or lower scan speed [10,15]. However, this approach has important limitations. Excessive energy input may promote keyhole instability, pore formation, evaporation, and melt-pool fluctuations, while the repeated thermal cycles inherent to PBF-LB may act as an in-situ heat treatment and induce non-equilibrium phase transformations during the build [10,16–19]. Moreover, volumetric energy density should be treated only as a first-order processing descriptor, since melt-pool behavior is governed by a coupled interaction among laser power, scan speed, hatch spacing, layer thickness, absorptivity, powder packing, and heat dissipation [18]. Therefore, maximizing nominal energy density is not necessarily the most effective route to chemical and microstructural homogenization.

An alternative strategy proposed in this study is first to use pre-alloyed master-alloy powders (PMAPs) with less dissimilar melting temperatures while maintaining the compositional freedom; and second to elaborate further the deposition process itself.

The use of PMAPs avoids the presence of pure refractory Nb particles in the powder bed and hence decreases the diffusion length required for complete chemical homogenization during deposition and subsequent heat treatment. Compared to strongly elemental blends, master-alloy powders also reduce local compositional extremes associated with pure Fe or Sn particles, while preserving compositional flexibility through controlled powder mixing. This approach is particularly attractive for Ti-Nb-Fe-Sn alloys, where several target compositions can be produced from a limited set of master-alloy powders rather than from a fully pre-alloyed powder batch for each alloy.

The PBF-LB process can also be adjusted for efficient homogenization without relying solely on higher laser power. A thicker powder layer on one hand reduces the nominal volumetric energy input for a fixed laser power, scan speed, and hatch spacing, but on the other hand may enhance homogeneity of distribution of powder particles of different pre-alloy, and further may alter the local balance between melt-pool size, cooling rate, and thermal gradient. By reducing the energy input, the use of thicker layers increases the risk of lack-of-fusion defects if melting continuity is insufficient. In this study it is therefore proposed to mitigate this disadvantage by additional layer remelting step. Additional layer remelting by laser beam without adding more powder is a useful complementary strategy, since it increases the cumulative laser-material interaction, promotes additional melt-pool mixing, hence homogeneity, and can reduce porosity without additional adverse increasing peak laser power [5,17,18]. In Ti alloys, where strong thermal gradients and epitaxial β growth often promote columnar grains, texture, and anisotropy, such control of the thermal path is particularly relevant [12,13,16].

Post-processing heat treatment is also essential for metastable β titanium alloys produced by PBF-LB. The as-built condition may contain residual stresses, dendritic or cellular segregation, and, most importantly, non-equilibrium phases such as α'/α'' or ω, depending on composition and thermal history [13,16]. These features can obscure the intrinsic effect of alloy composition and may compromise mechanical reliability. Heat treatment is therefore required to relieve residual stress, promote chemical homogenization, dissolve undesirable non-equilibrium phases, and establish a more representative β microstructure for property evaluation. Importantly, reducing the amount of pure Nb in the initial powder blend should lower the burden placed on heat treatment, potentially enabling homogenization under less severe conditions than would be required for elemental Nb-rich powder mixtures.

In this work, we evaluate the use of PMAPs, tailored PBF-LB processing, and subsequent heat-treatment for producing several homogeneous metastable β Ti-Nb-Fe-Sn alloys. The approach uses PMAPs mixed with commercially pure Ti to preserve compositional flexibility while avoiding pure refractory Nb in the feedstock. The PBF-LB route combines a relatively thick powder layer with layer remelting and subsequent heat treatment to promote melting continuity and solute redistribution without relying exclusively on excessive laser power. We selected four compositions from the Ti–Nb–Fe–Sn system explored previously for biomedical applications and successful additive manufacturing [3].

The main aim of this work is to establish whether a master-alloy powder blending strategy can be used to produce chemically homogeneous metastable β Ti-Nb-Fe-Sn alloys by PBF-LB while preserving compositional flexibility. Four alloys with progressively decreasing Fe and increasing Nb contents were fabricated and heat-treated to systematically evaluate the influence of composition on microstructure, grain-size evolution, elastic behavior, and mechanical properties.

Furthermore, thermodynamic calculations are used to determine whether solidification descriptors account for the non-monotonic grain-size evolution and to identify possible contributions from transient phase formation and solute partitioning during heat treatment. Hall-Petch and solid-solution strengthening analyses are used to distinguish grain-size effects from composition-dependent strengthening. The working hypothesis is that the master-alloy powder strategy enables the production of compositionally tunable Ti-Nb-Fe-Sn alloys, while their final microstructure and mechanical response are governed jointly by inherited solidification features and composition-dependent evolution during heat treatment.

## 2. Methods

### 2.1 Samples production

Four Ti-Nb-Fe-Sn alloy blends were designed with decreasing Fe content and increasing Nb content, while maintaining a constant Sn addition. The target powder-blend compositions were Ti-23Nb-3Fe-4Sn, Ti-26Nb-2Fe-4Sn, Ti-29Nb-1Fe-4Sn, and Ti-32Nb-4Sn (in wt.%), hereafter referred to as 23Nb-3Fe, 26Nb-2Fe, 29Nb-1Fe, and 32Nb-0Fe, respectively. These designations correspond to the nominal blend compositions and are used throughout the manuscript for consistency. These compositions were selected from the broader Ti–Nb–Fe–Sn alloy system previously investigated in Ref. [3].

The alloys were produced from powder blends containing pre-alloyed Ti-42Nb, Ti-20Nb-15Fe, and Ti-20Nb-20Sn master-alloy powders, together with commercially pure Ti Grade 2 powder for compositional adjustment. The Ti-42Nb powder was supplied by Taniobis GmbH, Germany, with a particle size distribution of 2–15 μm. The Ti-20Nb-15Fe and Ti-20Nb-20Sn powders were supplied by CAMEX, Czech Republic, with particle size distributions of 15–65 μm. Commercially pure Ti Grade 2 powder was supplied by AP&C, Canada, with a particle size distribution of 15–45 μm. All powders exhibited a predominantly spherical morphology. The target composition of each blend was calculated by mass balance from the composition and mass fraction of each powder. Powder blending was carried out in an eccentric mixer at 140 rpm for 2 h.

Samples were additively manufactured using an Omnisint-160 (Omnitek Technology) system equipped with a Yb:YAG fiber laser operating at a wavelength of 1070 nm. Cuboid samples with dimensions of 25 × 10 × 10 $mm^3$ were fabricated on a Ti-6Al-4V substrate under a dynamic argon atmosphere. During processing, the oxygen concentration was maintained below 500 ppm.

The processing parameters were selected based on previous optimization studies performed for Ti-Nb-Fe-Sn alloys [3,20]. The laser power, scanning speed, layer thickness, hatch spacing, and rotation angle between successive layers were set to 200 W, 100 mm/s, 70 μm, 80 μm, and 67°, respectively. An additional remelting step, using the same parameters, was applied to each layer to reduce porosity in the fabricated parts. This strategy was adopted as remelting can improve melt-pool stability and promote additional consolidation, thereby reducing lack-of-fusion defects and residual porosity in PBF-LB parts [5].

After fabrication, all samples were heat treated in an Ar atmosphere at 1000 °C for 2 h using a heating rate of 200 °C/h, followed by water quenching.

### 2.2 Chemical and microstructural characterization

Porosity was quantified by image analysis on longitudinal cross-sections of the printed samples using ImageJ software [21]. For each alloy, three regions larger than 4.5 × 5.5 mm² were analyzed.

Microstructural characterization was carried out by scanning electron microscopy (SEM) and electron backscatter diffraction (EBSD) using an Apreo 2S (ThermoFisher Scientific). Chemical analyses were performed by energy-dispersive X-ray spectroscopy (EDS) using an Auriga Compact microscope (ZEISS) equipped with an EDS detector. Samples for SEM, EDS, and EBSD analyses were prepared using conventional metallographic procedures, including grinding and polishing, followed by final vibratory polishing using a VibroMet 2 system (Buehler). EBSD maps were acquired over large areas of approximately 7,000 × 5,000 μm² using a step size of 8 μm to improve the statistical reliability of the grain-size and texture analyses.

Oxygen and nitrogen contents were measured by carrier gas hot extraction using a Bruker Galileo G8 analyzer.

Phase identification was performed by X-ray diffraction using a Rigaku Rapid II diffractometer equipped with a Mo-Kα radiation source operating at 50 kV and 40 mA.

Diffraction patterns were acquired in reflection mode using a curved two-dimensional detector and a graphite monochromator. The two-dimensional diffraction data were subsequently integrated into one-dimensional diffraction patterns. Instrumental peak-profile parameters and the angular zero offset were calibrated using a NIST SRM 660c $LaB_6$ standard measured under identical conditions. The lattice parameters were subsequently obtained by Rietveld refinement using the Profex software [22].

Grain size analysis was performed from the EBSD maps using MTEX [23]. The EBSD data were first filtered based on the confidence index threshold of 0.1 and then smoothed using a median filter. Grains were reconstructed using a critical misorientation angle of 5° and a minimum grain size threshold of 10 pixels. Grains intersecting the map boundaries were excluded from the analysis. These criteria were adopted following the general requirements of ISO 13067 and ASTM E2627, as discussed by [24], including the use of a suitable EBSD step size. Due to the large grain size observed in the 26Nb-2Fe and 29Nb-1Fe alloys, the recommended number of 500 grains could not be reached for these alloys. Nevertheless, more than 400 grains were analyzed in each case, which was considered sufficient to provide a statistically meaningful comparison among the compositions investigated.

### 2.3. Thermodynamic calculations

Thermodynamic calculations were carried out on Thermo-Calc software using the TCTI4 database to support the interpretation of phase stability, solidification behaviour, and β grain-size evolution. The calculations were carried out using the measured chemical compositions of the heat-treated alloys rather than the nominal composition. Equilibrium calculations were performed over the temperature range relevant to solidification and heat treatment to identify stable phase fields, phase fractions, and solute partitioning among the predicted phases. Scheil simulations [25,26] were also performed to evaluate non-equilibrium solidification paths and to estimate the freezing interval associated with each composition.

### 2. 4 Mechanical properties

Microhardness measurements were performed using a Qness Q10 microhardness tester employing a load of 500 g and a dwell time of 10 s. For each sample, hardness maps were constructed over an area of 4.8 × 3 $mm^2$, with indentations spaced by 600 μm, totaling 54 indentations.

Tensile tests were carried out in triplicate using an Instron 5882 universal testing machine equipped with a video extensometer. The tests were performed at room temperature with an initial strain rate of $1 \times 10^{-3} s^{-1}$. The gauge length of the samples was 5 mm, width 2 mm and thickness 1 mm.

Elastic constants were measured by resonant ultrasound spectroscopy (RUS) [27], using a contactless laser-based system [28]. Rectangular parallelepiped specimens with approximate dimensions of 5 x 4.5 x 4 $mm^3$ were cut from the PBF-LB samples and their largest faces were polished using silica suspension (Struers OP-S NonDry, 0.25 μm) to maximize laser reflectance. During the measurement, short infrared laser pulses (Quantel ULTRA 1064 nm Nd:YAG laser) were used to excite free mechanical vibrations in the specimen, while the surface response was recorded by a scanning laser interferometer (Polytec MSA-600 Micro System Analyzer). Measurements were conducted in low-pressure nitrogen atmosphere with precise temperature control at 20 °C. The resonant frequencies and vibration modes were then compared with calculated eigenfrequencies, and the elastic constants were obtained by inverse fitting [29]. As the PBF-LB rotation angle between successive layers was 67°, which does not lead to a periodic pattern over a large number of layers, and as the RUS samples were made of around 70 total layers, the samples were assumed to be transversely isotropic.

## 3. Results

### 3.1 Microstructural characterization

Figure 1 shows SEM-BSE micrographs of the heat-treated 23Nb-3Fe, 26Nb-2Fe, 29Nb-1Fe and 32Nb-0Fe samples. After heat treatment, no residual contrast associated with unmelted particles, cellular/dendritic segregation, or melt-pool boundaries were observed within the resolution of SEM-BSE. Image analysis revealed low porosity levels of 0.37 (± 0.12) %, 0.27 (± 0.02) %, 0.19 (± 0.14) %, and 0.24 (± 0.10) % for the 23Nb-3Fe, 26Nb-2Fe, 29Nb-1Fe, 32Nb-0Fe alloys, respectively.

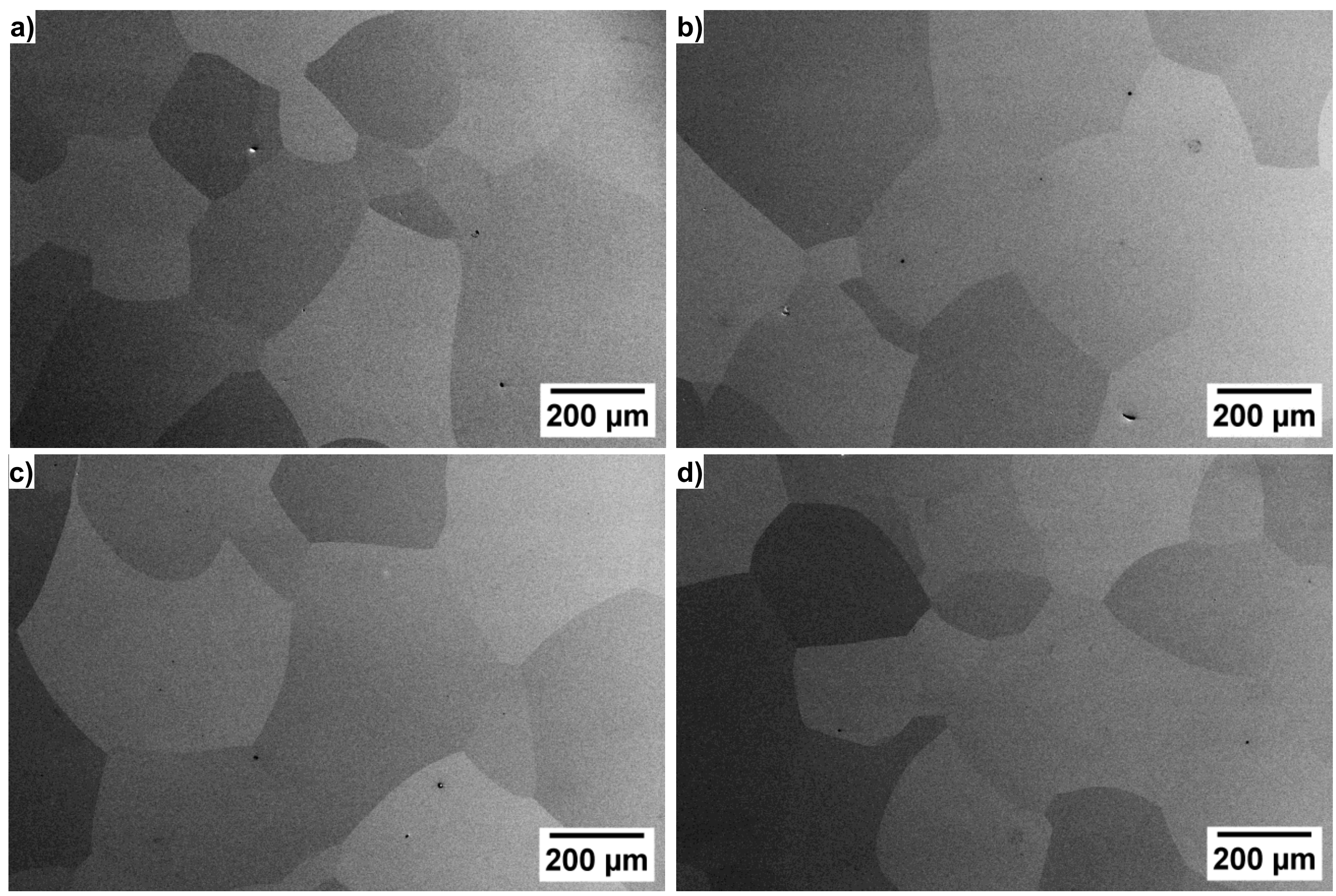

Figure 1. SEM-BSE micrographs of the heat-treated (a) 23Nb-3Fe, (b) 26Nb-2Fe, (c) 29Nb-1Fe, and (d) 32Nb-0Fe alloys.

Table 1 summarizes the measured chemical compositions of the investigated alloys. The compositions follow the intended trend of increasing Nb and decreasing Fe contents, while Sn and the interstitial O and N contents remain nearly constant across the alloy series. The low EDS standard deviations indicate a homogeneous distribution of Nb, Fe, and Sn after heat treatment.

Table 1. Measured chemical composition of the heat-treated Ti-Nb-Fe-Sn alloys.

| Alloy | Composition (wt. %) | | | | | |
|---|---|---|---|---|---|---|
| | Ti | Nb | Fe | Sn | O | N |
| Ti-32Nb-4Sn | Balance | 33.7 ± 0.5 | 0 | 4.6 ± 0.1 | 0.216±0.032 | 0.037±0.004 |
| Ti-29Nb-1Fe-4Sn | | 27.8 ± 0.1 | 1.1 ± 0.1 | 3.3 ± 0.1 | 0.236±0.018 | 0.036±0.005 |
| Ti-26Nb-2Fe-4Sn | | 26.3 ± 0.2 | 1.6 ± 0.1 | 3.6 ± 0.1 | 0.202±0.011 | 0.035±0.003 |
| Ti-23Nb-3Fe-4Sn | | 23.4 ± 0.3 | 3.0 ± 0.2 | 4.8 ± 0.2 | 0.218±0.012 | 0.033±0.010 |

Figure 2 shows the XRD patterns of the heat-treated alloys. In all cases, only reflections corresponding to the β phase were detected, with no evidence of α, α', α'' or ω peaks within the detection limits of the technique. The calculated β lattice parameters were 3.261 Å, 3.265 Å, 3.267 Å, and 3.274 Å for the 23Nb-3Fe, 26Nb-2Fe, 29Nb-1Fe, and 32Nb-0Fe alloys, respectively. A weak positive dependence of the β lattice parameter on Nb content has previously been reported for binary Ti–Nb alloys [30,31], whereas increasing Fe content has been shown to contract the β lattice in binary Ti–Fe alloys [32]. Therefore, the present trend is consistent with the combined increase in Nb content and decrease in Fe content. The narrower variation in Sn content makes its individual contribution less evident in the present dataset.

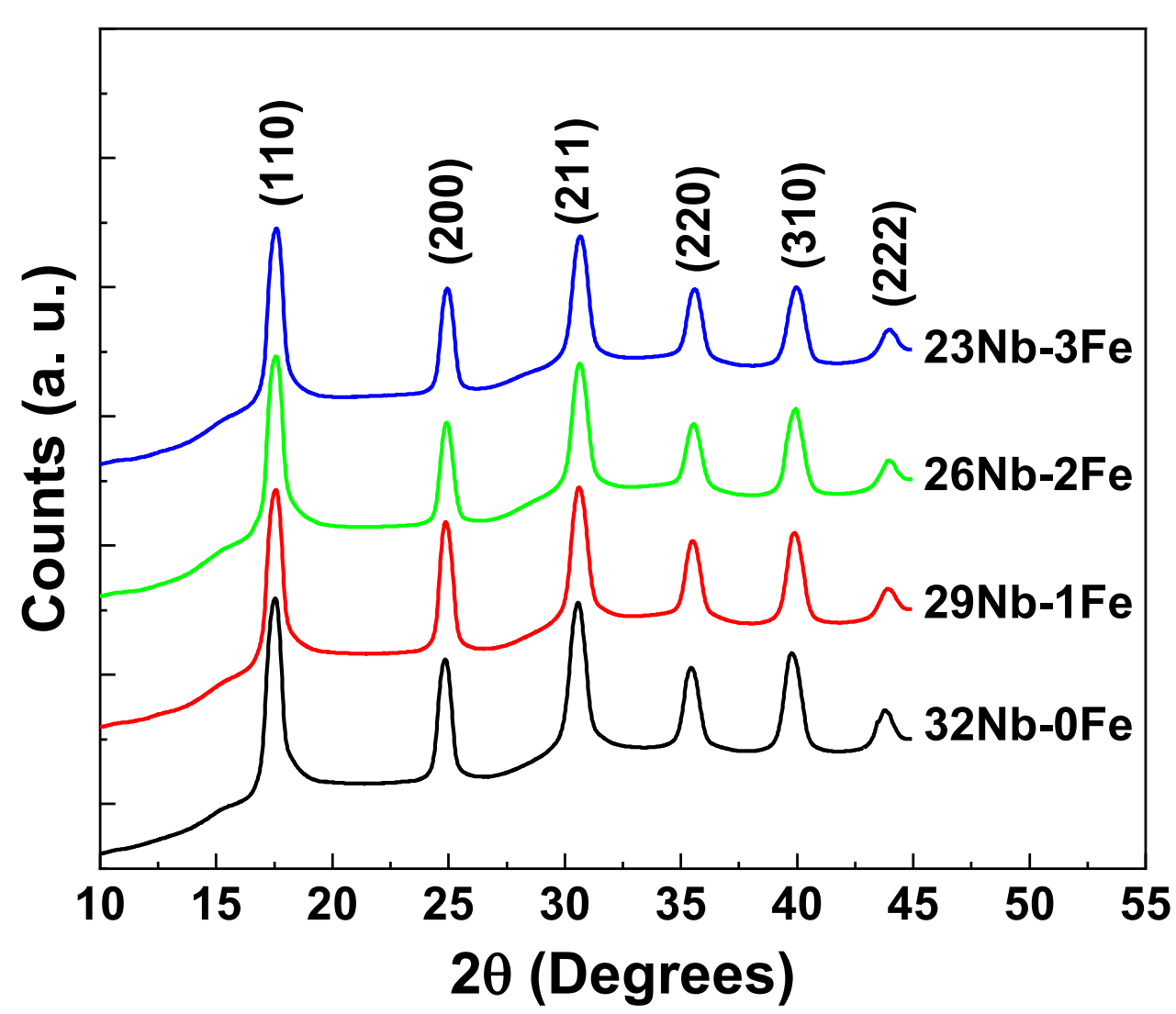

Figure 2. XRD patterns of the heat-treated single β-phase 23Nb-3Fe, 26Nb-2Fe, 29Nb-1Fe, and 32Nb-0Fe alloys. Counts were plotted logarithmically to highlight that there are no low intensity reflections.

Figure 3 presents the EBSD inverse pole figure maps (IPF-Z) overlaid with the grain boundaries, together with the corresponding grain size distributions in terms of equivalent diameter. All samples exhibited predominantly equiaxed β grains after heat treatment. The area-weighted mean β grain sizes, $D$, were 298 (± 123) μm, 419 (± 146) μm, 467 (± 174) μm, and 359 (± 140) μm for the 23Nb-3Fe, 26Nb-2Fe, 29Nb-1Fe, and 32Nb-0Fe alloys, respectively. Grain size exhibited a non-monotonic dependence on composition, with the finest grains observed in the 23Nb-3Fe alloy and the coarsest in the 29Nb-1Fe alloy.

The average area-weighted aspect ratios (AR) were 1.47, 1.47, 1.57, and 1.58 for the 23Nb-3Fe, 26Nb-2Fe, 29Nb-1Fe, and 32Nb-0Fe alloys, respectively. These values are below 2, indicating predominantly equiaxed grains [33]. Qualitatively, the IPF-Z maps also indicate weak crystallographic texture in all alloys.

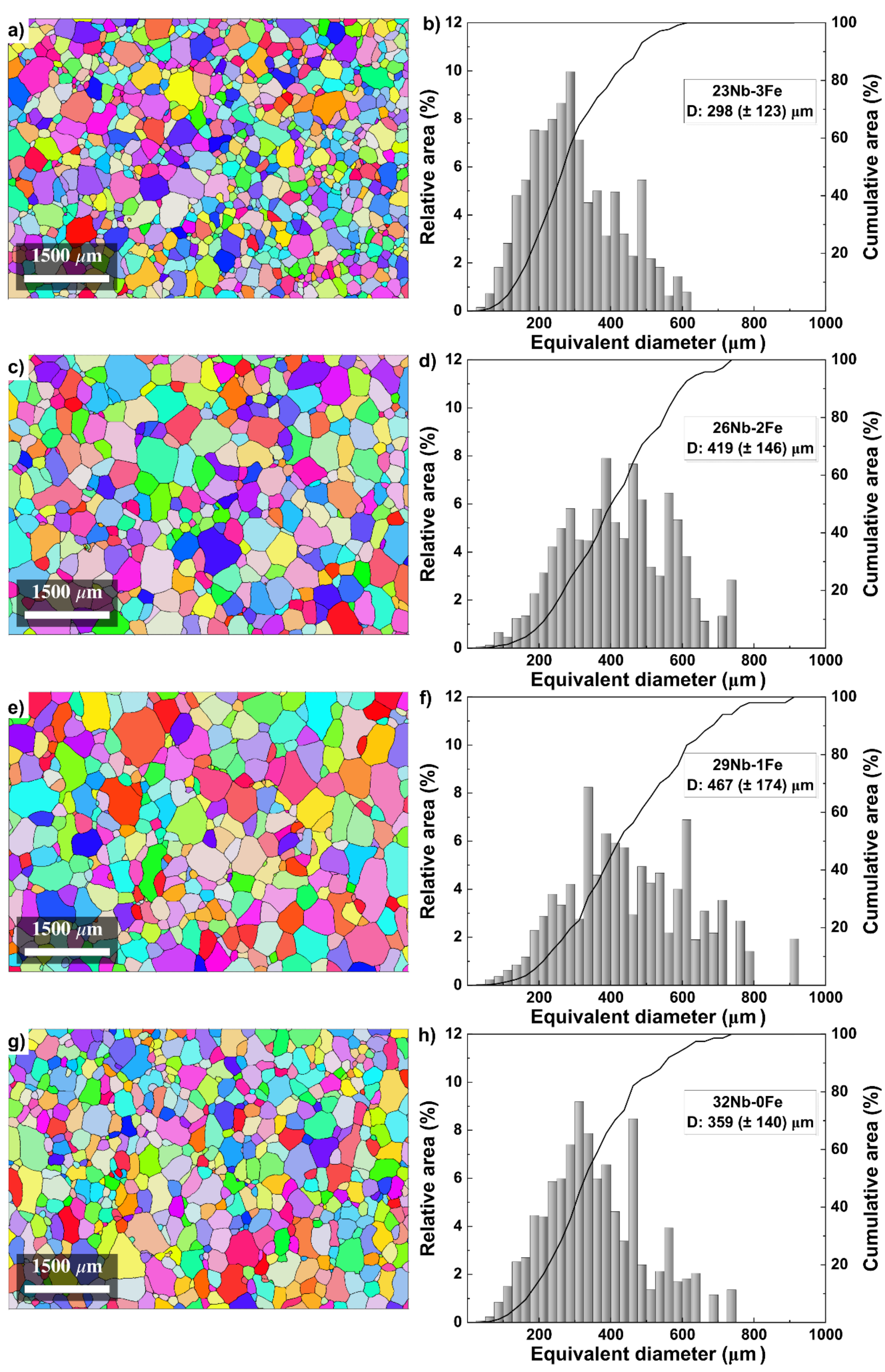

Figure 3. EBSD analysis of the heat-treated alloys: IPF-Z maps and β grain-size distributions for (a,b) 23Nb-3Fe, (c,d) 26Nb-2Fe, (e,f) 29Nb-1Fe, and (g,h) 32Nb-0Fe alloys.

Figure 4 shows the (001), (011), and (111) pole figures of the heat-treated alloys. All samples exhibited low texture intensity, with maximum values of approximately 1.8 MRD. This confirms that the heat-treated microstructures were weakly textured and close to random. Together with the equiaxed grain morphology and homogeneous SEM-BSE contrast, these results indicate that the combination of the selected PBF-LB strategy and post-processing heat treatment promoted dense, chemically homogeneous, and weakly textured β microstructures.

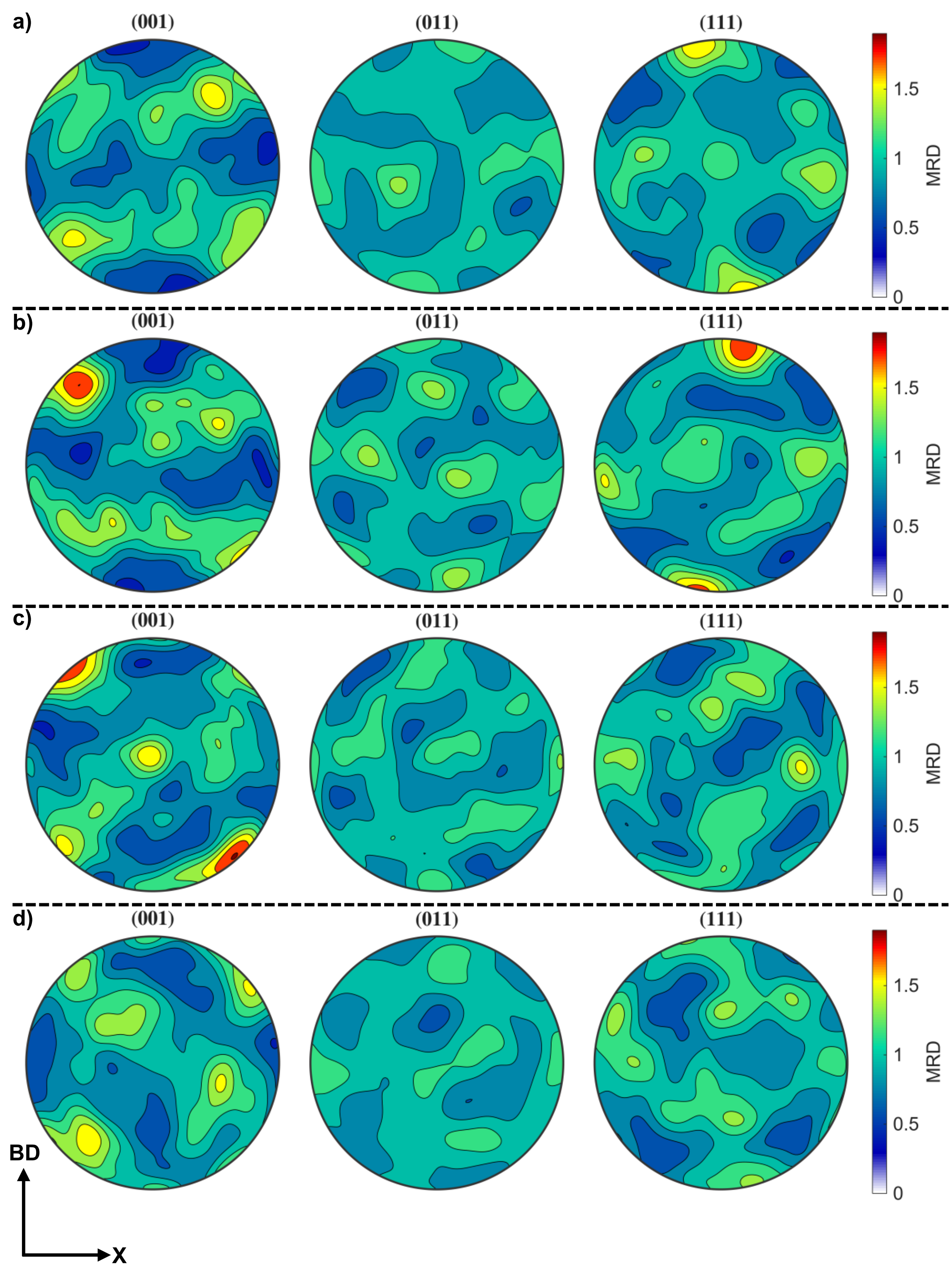

Figure 4. (001), (011), and (111) pole figures of the heat-treated (a) 23Nb-3Fe, (b) 26Nb-2Fe, (c) 29Nb-1Fe, and (d) 32Nb-0Fe alloys.

### 3.2 Mechanical properties

Figure 5 presents the hardness maps of the heat-treated alloys. The average hardness values were 268 (± 4) HV, 249 (± 6) HV, 241 (± 4) HV, and 224 (± 4) HV for the 23Nb-3Fe, 26Nb-2Fe, 29Nb-1Fe, and 32Nb-0Fe samples, respectively, reflecting the increase in Nb content and decrease in Fe content. Since the Sn content varied only slightly among the alloys, its individual contribution to hardness could not be assessed. The observed trend is consistent with the expected stronger strengthening effect of Fe in β-Ti alloys. Fe has been reported to exhibit a higher Hall-Petch strengthening coefficient ($k_i$) than Nb, with values of approximately 1589 MPa·μm$^{0.5}$ for Fe and 1051 MPa·μm$^{0.5}$ for Nb [34]. Although the corresponding $k_i$ for Sn has not been reported, its contribution is generally considered negligible because of its relatively low shear modulus (~18 GPa) [24-26].

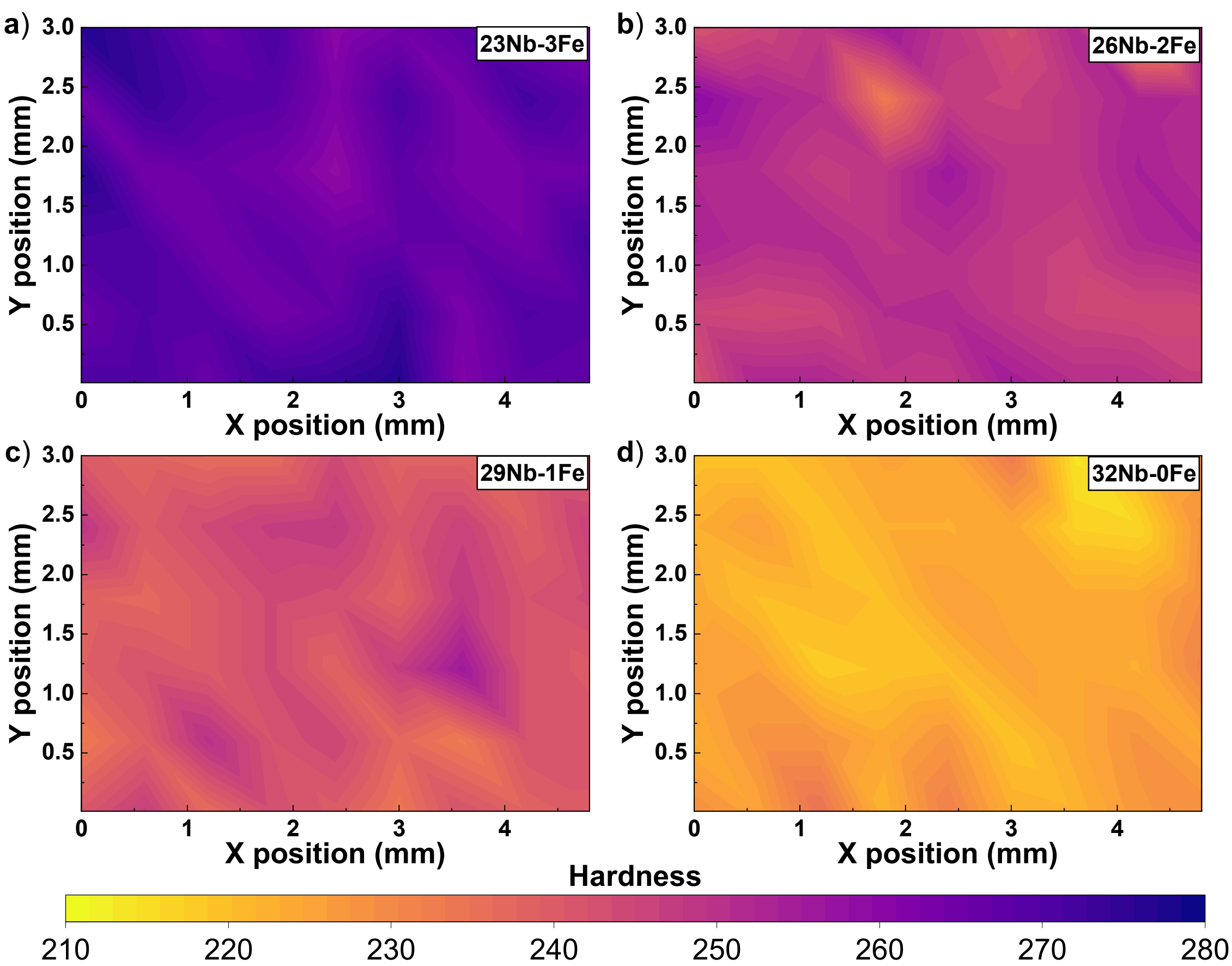

Figure 5. Hardness maps of the heat-treated (a) 23Nb-3Fe, (b) 26Nb-2Fe, (c) 29Nb-1Fe, and (d) 32Nb-0Fe alloys.

Figure 6 presents the tensile engineering stress-strain curves from three specimens for each alloy, illustrating the reproducibility and scatter in mechanical response. The average yield strengths $\sigma_y$ were (691 ± 13) MPa, (615 ± 11) MPa, (606

± 13) MPa, and (468 ± 14) MPa for the 23Nb-3Fe, 26Nb-2Fe, 29Nb-1Fe, and 32Nb-0Fe alloys, respectively. Consistent with the hardness results, $\sigma_y$ decreased with decreasing Fe and increasing Nb contents. The contributions of the individual strengthening mechanisms are discussed in the following section. The 26Nb-2Fe and 29Nb-1Fe alloys exhibited the largest scatter in ultimate tensile strength and ductility, which coincided with their larger β grain sizes. Based on the measured *D* values and the tensile specimen cross-sectional area of approximately 2 mm², only approximately 12-15 grains are expected within the gauge sections of these alloys, compared with at least 20 and 30 grains for 32Nb-0Fe and 23Nb-3Fe, respectively. The greater scatter in the coarser-grained alloys is therefore attributed to representative volume effects and the stronger influence of local crystallographic orientation on the tensile response.

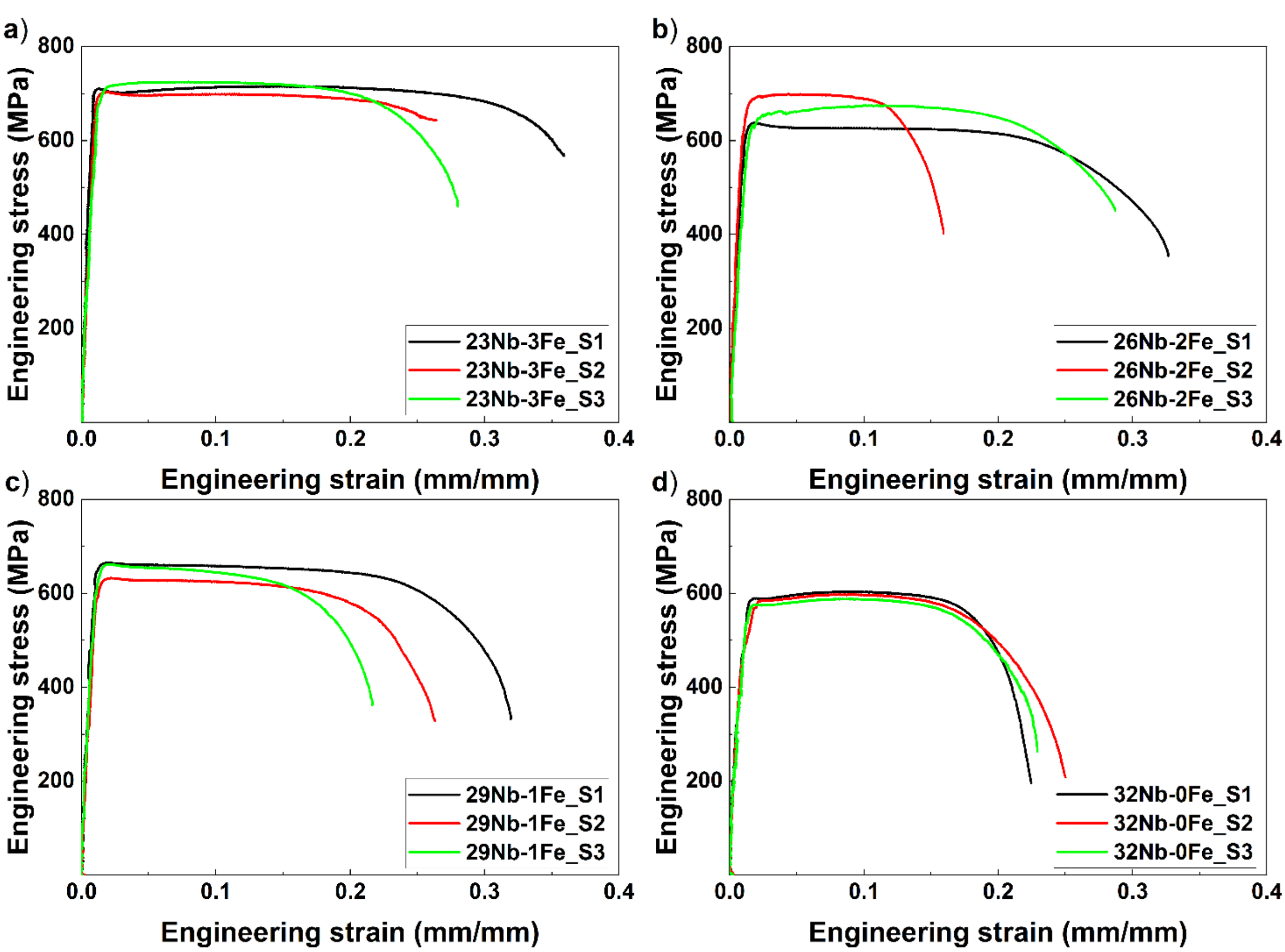

Figure 6. Engineering stress-strain curves of the heat-treated (a) 23Nb-3Fe, (b) 26Nb-2Fe, (c) 29Nb-1Fe, and (d) 32Nb-0Fe alloys.

Figure 7 presents the elastic properties measured by resonant ultrasound spectroscopy. Panel (a) shows the directional Young's modulus, $E$, relative to the horizontal plane. Panel (b) compares the in-plane modulus with the minimum and maximum moduli obtained for out-of-plane directions. The RUS results were described assuming transverse isotropy, meaning that the elastic response is identical in all directions within the horizontal plane but may differ along the build direction. The slight anisotropy observed in the samples likely arises from the combined effects of the weak residual crystallographic texture and the composition-dependent elastic constants of their respective crystal structures.

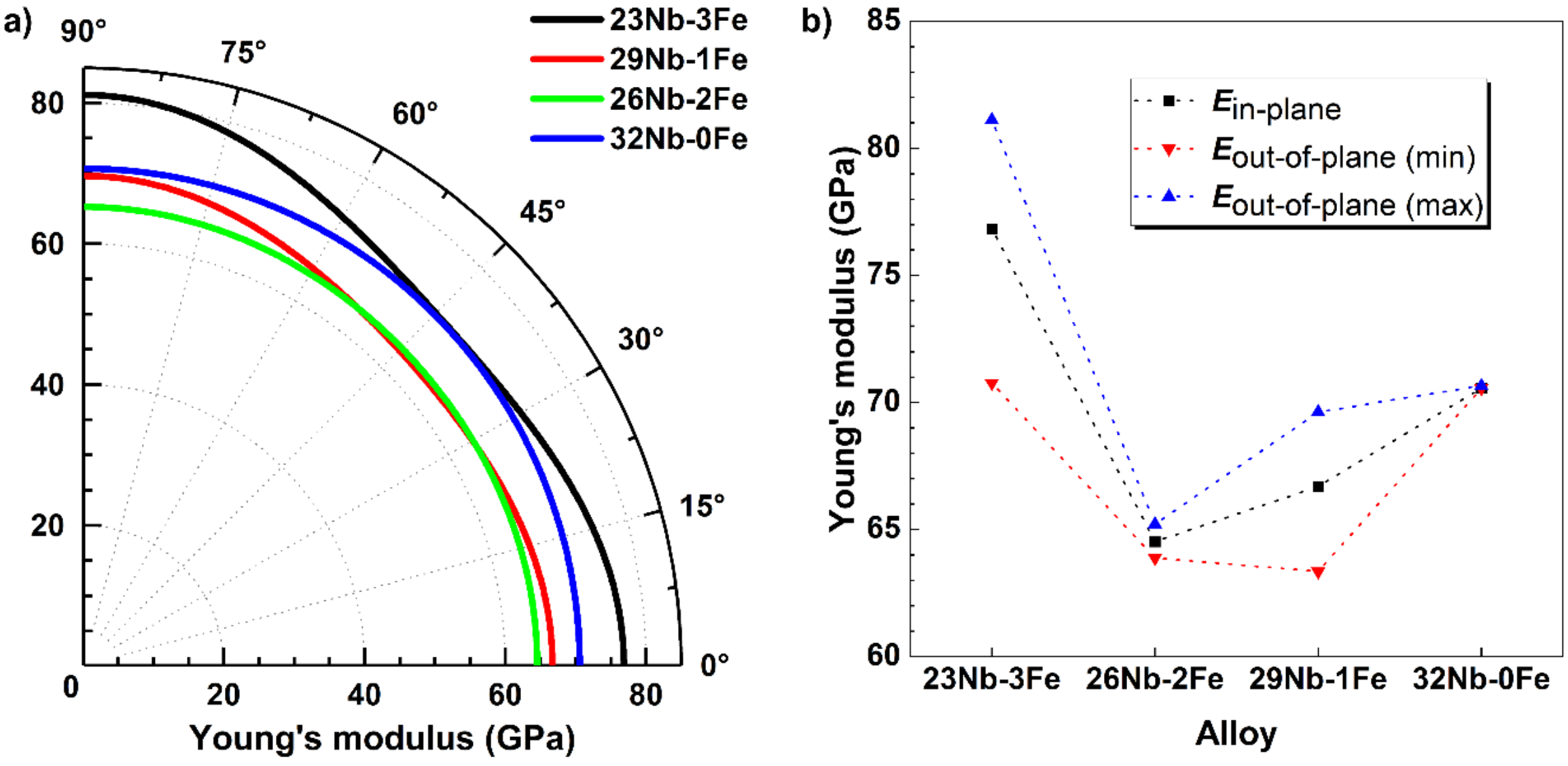

Figure 7. Directional Young's modulus of the 23Nb-3Fe, 26Nb-2Fe, 29Nb-1Fe, and 32Nb-0Fe alloys: (a) angular distribution of the Young's modulus in transverse isotropic symmetry, where 0° angle corresponds to the horizontal plane and 90° angle corresponds to the build direction. (b) Comparison between the in-plane Young's modulus ($E_{in\text{-}plane}$) and the minimum and maximum out-of-plane values ($E_{out\text{-}of\text{-}plane}$).

All investigated alloys exhibited relatively low Young's moduli, ranging from approximately 63 to 81 GPa depending on composition and crystallographic direction. Although the elastic moduli differed among the investigated compositions, no monotonic dependence on Nb or Fe content was observed. For all samples, the build direction (90°) was the stiffest direction with the highest $E$ values, and the lowest $E$ values were observed near the diagonal direction (close to the 45° angle in all studied samples). Nevertheless, the level of anisotropy differs among the samples. The 26Nb-2Fe and 32Nb-0Fe samples can be assumed isotropic, as its differences in the angular distribution of Young's moduli are much lower than the measurement uncertainty (approximately ± 2 GPa). In other samples, the elastic anisotropy corresponds well to the texture observed by EBSD. The build direction exhibits a relatively high (111) texture and consequently the highest Young's modulus values, as (111) is the stiffest direction in beta Ti alloys, whereas (001) is the softest [35,36].

## 4. Discussion

### 4.1 Microstructural homogenization and β grain-size evolution

The heat-treated alloys exhibited low porosity, homogeneous SEM-BSE contrast, and single-β phase constitution by XRD. The master-alloy powder strategy combined with advanced PBF-LB processing and homogenization heat treatment therefore successfully produced chemically homogeneous β Ti-Nb-Fe-Sn. The proposed and realized strategy effectively suppressed the heterogeneities typically associated with PBF-LB processing of Ti-Nb-based alloys. No evidence of unmelted Nb-rich particles, cellular/dendritic segregation, or melt-pool boundaries was observed after heat treatment. The hardness maps revealed only minor spatial variations within each composition. The highest standard deviation was 6 HV for the 26Nb-2Fe alloy, indicating the high degree of homogeneity across the analyzed regions.

The β grain-size sequence was 23Nb-3Fe < 32Nb-0Fe < 26Nb-2Fe < 29Nb-1Fe, corresponding to area-weighted mean β grain sizes of approximately 298, 359, 419, and 467 μm, respectively. This non-monotonic trend indicates that the final heat-treated grain size cannot be explained solely by Nb or Fe content but instead reflects the combined effects of the inherited solidification structure, transient phase evolution during heating, solute partitioning, and β grain-boundary mobility during the 1000 °C heat treatment.

To assess whether this trend originates during solidification, thermodynamic calculations were performed to estimate the growth-restriction factor (*Q)* and the constitutional-undercooling parameter (*P)* using the measured compositions.

The growth-restriction factor quantifies the development of constitutional undercooling ahead of the solid-liquid interface during the early stages of solidification. In this work, *Q* was calculated from the initial slope of the Scheil solidification curve, using the multicomponent expression [37]:

$$Q = \left(\frac{\partial(\Delta T_{CS})}{\partial f_s}\right)_{f_s \to 0} \quad \text{Eq. 1}$$

where $f_s$ is the solid fraction and $\Delta T_{CS}$ is the undercooling relative to the liquidus temperature. The derivative was determined by linear regression over the initial Scheil region, 0.005 ≤ $f_s$ ≤ 0.05, where growth restriction is expected to play the most significant role in solute enrichment ahead of the solid-liquid interface.

The *P* parameter was treated following the approach used by Brooke et al. [38] for titanium alloys, in which *P* is approximated by the equilibrium freezing range:

$$P = T_L^{eq} - T_S^{eq} \quad \text{Eq. 2}$$

where $T_L^{eq}$ and $T_S^{eq}$ are the equilibrium liquidus and solidus temperatures, respectively. This definition is provided because, in the present treatment, *P* is not a fitted parameter, but a thermodynamic descriptor obtained from equilibrium calculations.

Table 2 summarizes the calculated descriptors together with the measured β grain size after heat treatment. Since *Q* is inversely proportional to grain size during solidification [39], the addition of Fe combined with the reduction in Nb is expected to enhance grain refinement by substantially increasing the Scheil freezing interval from the Fe-free 32Nb-0Fe alloy to the Fe-containing alloys. The largest increase occurs between the Fe-free 32Nb-0Fe and 29Nb-1Fe, where the Scheil freezing interval rises from approximately 203 K to 649 K. This change is primarily attributed to the addition of ~ 1.1 wt.% Fe, since Fe strongly segregates to the remaining liquid and produces a long low-temperature tail in the Scheil model. Further increases in the Fe content produce smaller changes in the Scheil freezing interval.

For the Fe-containing alloys, 23Nb-3Fe, 26Nb-2Fe, and 29Nb-1Fe, the β grain size increases as *Q* and *P* decrease, consistent with weaker growth restriction and reduced constitutional undercooling leading to coarser inherited β grains. However, the Fe-free 32Nb-0Fe alloy does not follow this simple solidification-controlled trend. Despite exhibiting the highest *Q* and lowest *P*, it develops an intermediate grain size rather than the finest or coarsest structure, respectively.

Table 2. Solidification descriptors calculated from Thermo-Calc Scheil and equilibrium simulations.

| Alloy | β grain size (μm) | Scheil freezing interval (K) | *P* (K) | *Q* (K) |
|---|---|---|---|---|
| 23Nb-3Fe | 298 | 599 | 125 | 82 |
| 26Nb-2Fe | 419 | 637 | 90 | 70 |
| 29Nb-1Fe | 467 | 649 | 77 | 68 |
| 32Nb-0Fe | 359 | 203 | 66 | 90 |

A larger freezing range is generally associated with increased undercooling and nucleation rates [39]. Although the 32Nb-0Fe alloy has a much smaller Scheil freezing interval and P than the Fe-containing alloys, it does not exhibit the largest grain size after heat treatment. Therefore, the calculated solidification descriptors explain part of grain-size evolution, particularly among the Fe-containing compositions, but are insufficient to predict the final β grain size sequence after heat treatment. This mismatch is expected because the measured grain sizes correspond to the post-heat-treatment β microstructure. The heating ramp and the subsequent 2 h hold at 1000 °C provide substantial thermal exposure for recovery, possible static recrystallization, phase dissolution, and grain growth. Therefore, the final grain size results from the combined effects of the inherited as-built structure and the subsequent thermal evolution during heat treatment.

To identify possible phase transformations that can influence the grain growth during heating and the 1000 °C hold, equilibrium calculations were performed, as shown in Figure 8. All alloys are predicted to be single-phase BCC β solid solutions at the heat-treatment temperature. Upon cooling, however, the BCC phase field separates into compositionally distinct BCC phase instances and an HCP phase. The BCC-A2#1, BCC-A2#2, and BCC-A2#3 labels therefore correspond to compositionally distinct BCC phase predictions from the thermodynamic database and do not necessarily represent experimentally observed phases. BCC-A2#1 corresponds to a Ti- and Sn-rich BCC phase associated with the subsequent formation of the HCP phase, whereas BCC-A2#2 is comparatively enriched in Nb and Fe. In the Fe-containing alloys, the calculations predict that BCC-A2#2 undergoes further compositional separation at approximately 500 °C, producing an Nb-rich BCC-A2#3 instance and an ordered Fe-rich BCC-B2 phase. The latter contains approximately 42–44 wt.% Fe and is therefore compositionally consistent with a TiFe-like intermetallic phase containing Nb in solid solution.

The maximum equilibrium fraction predicted for this Fe-rich phase decreases systematically with decreasing Fe content, from 5.46 vol.% in 23Nb-3Fe to 2.87 vol.% in 26Nb-2Fe and 1.96 vol.% in 29Nb-1Fe. The Fe-rich phase is predicted to be stable only at relatively low temperatures, dissolving in equilibrium near 473-506 °C depending on composition. Therefore, it cannot provide Zener pinning during the two-hour hold at 1000 °C, however, it may act as a transient phase during the heating ramp. At a heating rate of 200 °C/h, the alloys remain within the temperature range of Fe-rich stability for 20-30 min, followed by ~1 h in the HCP + BCC region before entering the single-β field. This thermal path provides a potential window for transient TiFe-like precipitation, subsequent dissolution, and β nucleation during heating, assuming sufficiently rapid transformation kinetics.

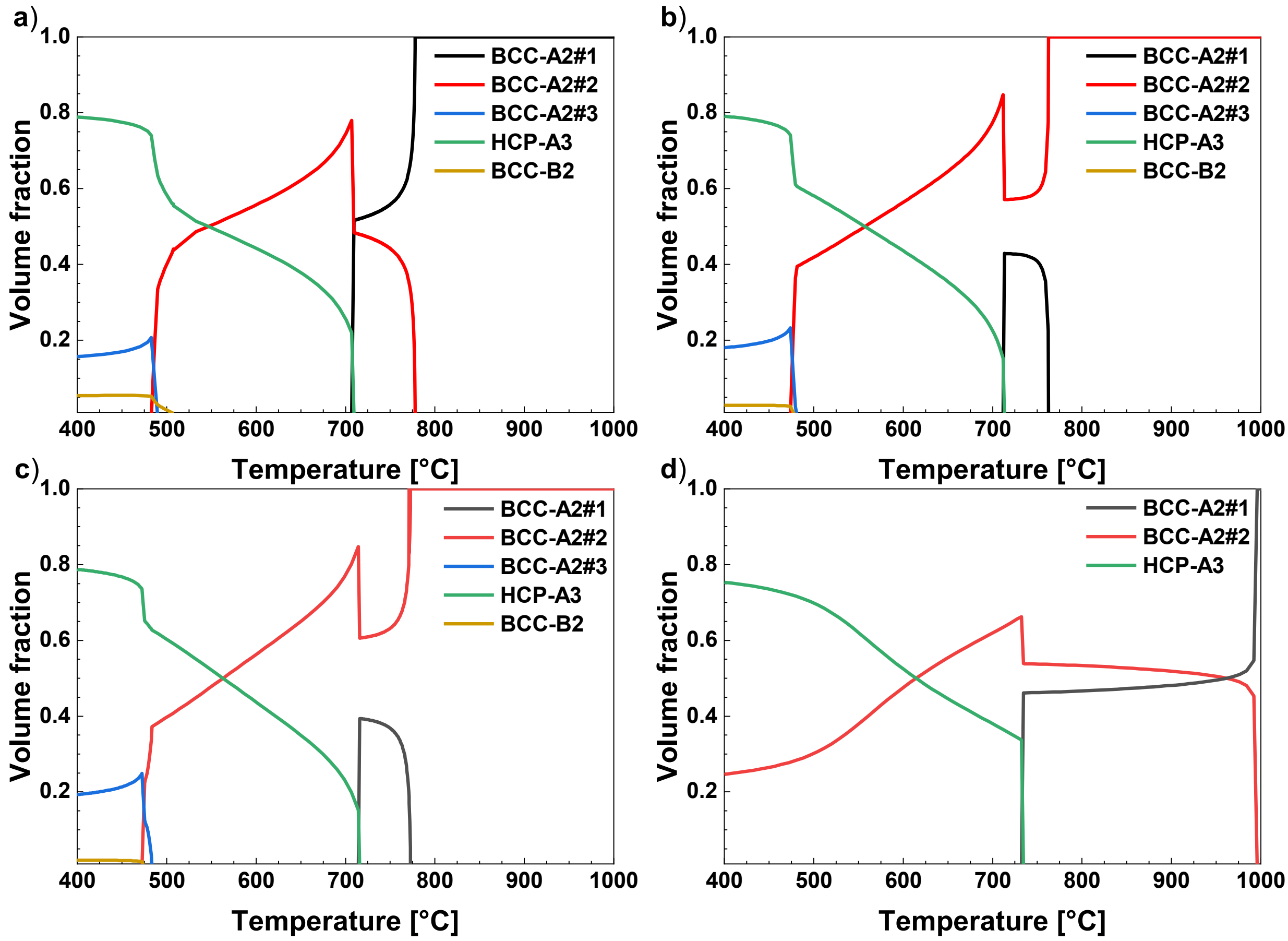

Figure 8. Equilibrium phase-fraction calculations for (a) 23Nb-3Fe, (b) 26Nb-2Fe, (c) 29Nb-1Fe, and (d) 32Nb-0Fe.

The mechanism proposed by Lee et al. [40] for Ti-Al-Fe-Sn alloys provides a useful framework for interpreting this behavior, although the terminology must be adapted to the present system. They report that adding Sn to the Ti-Al-Fe system promoted a grain refinement during hot rolling through enhanced TiFe stability and solute drag mechanisms. The solute-drag acts as barriers to the grain growth and TiFe acts as α-phase nucleation sites, which transforms into β phase during heating. Although our EDS-measured Sn content has varied just slightly compared to the nominal composition, when the average content is compared between samples, it can be more than 1.5 wt.%. Interestingly, the trend observed in the β grain sizes follows the same order as the measured Sn content.

In the present additively manufactured alloys, no thermomechanical deformation occurs during the heat treatment, as performed by Lee et al. [40]. However, the high residual stresses, dislocation density, subgrain structure, and chemical segregation inherited from additive manufacturing may provide stored energy analogous to that introduced by thermomechanical processing. Therefore, the relevant processes involved here should include recovery, possible static recrystallization, phase transformation, and grain growth. During slow heating, transient TiFe-like precipitation may interact with this stored energy by locally retarding boundary migration, delaying recovery, or providing Fe-enriched interfaces that promote β nucleation during the HCP-to-BCC transformation, similarly to the Lee et al. [40] report.

The role of Sn is also likely to be more complex than simply increasing the TiFe fraction. Owing to its large atomic radius and mass relative to Ti, Sn can produce solute-drag effects in titanium alloys [40,41]. The observed decrease in β grain size with increasing Sn content suggests that Sn may reduce β grain-boundary mobility through solute drag and/or through transient chemical partitioning during heating. For the Fe-free 32Nb-0Fe alloy, the intermediate final β grain size cannot be attributed to TiFe-assisted refinement, as no Fe-rich TiFe-like is predicted. Instead, the equilibrium calculations suggest a distinct mechanism associated with Nb/Sn partitioning within the β-phase field (Figure 8d).

During heating, the 32Nb-0Fe alloy is predicted to pass through a broad two-BCC region (~735 °C to 996 °C) before reaching the single-β field at the heat-treatment temperature. Within this range, the two BCC phases exhibit strong partitioning, with one BCC phase enriched in Nb and the other enriched in Sn. For instance, at ~ 800 °C, the Nb-rich BCC phase contains ~ 47.6 wt.% Nb and only ~0.5 wt.% Sn, whereas the complementary Ti/Sn -rich BCC phase contains ~ 15.5 wt.% Nb and ~10 wt.% Sn.

This behavior indicates a strong thermodynamic tendency for transient Nb/Sn chemical heterogeneity during heating.

Although the 32Nb-0Fe alloy is predicted to be fully β at 1000 °C, complete chemical homogenization is unlikely instantaneously because Nb and Sn diffuse slowly as substitutional solutes in β-Ti. Residual Nb/Sn-enriched regions or transient BCC/BCC interfaces may therefore reduce the effective β grain-boundary mobility during the early stages of the 1000 °C hold through solute drag. This provides a possible explanation for why the 32Nb-0Fe alloy does not exhibit the largest β grain size after heat treatment, despite the absence of an Fe-rich TiFe-like phase.

Overall, the final β grain size appears to result from competing mechanisms. In the Fe-containing alloys, the finer grain size of 23Nb-3Fe alloys may be related to the larger predicted fraction of transient Fe-rich TiFe-like phase during heating, which could retard recovery or promote additional β nucleation during phase transformation. In contrast, grain growth in 32Nb-0Fe may be limited by Nb/Sn partitioning and solute-drag effects associated with the transient two-BCC region below the single-β one. Therefore, the final β grain size after heat treatment reflects the combined effects of inherited solidification structure, transient phase evolution, solute partitioning, and β grain-boundary mobility during the 1000 °C hold.

### 4.2 Strengthening mechanisms

The hardness and yield strength decreased systematically from 23Nb-3Fe to 32Nb-0Fe, following the reduction in Fe content and the increase in Nb content.

To interpret the strengthening response of the heat-treated Ti-Nb-Fe-Sn alloys, some common strengthening mechanisms will be excluded as they do not play a major role. XRD and SEM-BSE showed that all alloys were fully β-BCC within the detection limits of the techniques, with no evidence of α, ω, or intermetallic precipitates after heat treatment. Therefore, precipitation strengthening is not expected to contribute to the measured yield strength.

Dislocation strengthening is also not expected to be the controlling contribution in the heat-treated condition. The 1000 °C/2 h treatment produced coarse, predominantly equiaxed β grains, indicating extensive boundary migration and a substantial reduction of the stored-energy state inherited from additive manufacturing. During annealing, recovery reduces stored energy through dislocation annihilation and rearrangement into lower-energy configurations, while recrystallization and subsequent grain growth further replace defect-rich regions by lower-defect grains [42]. This interpretation is also consistent with strengthening analyses reported for β-Ti alloys. Zhao et al. noted that rapidly solidified Ti-Fe-Sn-Nb alloys may retain dislocation densities on the order of $10^{15}$ $m^{-2}$, whereas solution-treated Ti alloys reported in the literature are typically below $10^{14}$ $m^{-2}$ [43,44]. Therefore, in the present heat-treated alloys, dislocation strengthening is likely a residual contribution rather than the main factor controlling the composition-dependent mechanical response.

The grain boundary strengthening contribution was calculated using the Hall-Petch relationship [45]:

$$\sigma_{HP} = k_y \cdot D^{-1/2} \qquad \text{Eq. 3}$$

where $k_Y$ is the Hall-Petch coefficient obtained as $k_Y = k_{Ti} + \sum_i k_i X_i$, in which the $k_i$ values for each element were taken from [34] and the $X_i$ is the atomic fraction of the alloying elements. Using the EDS-measured atomic fractions, the calculated Hall-Petch contributions were 55, 46, 44, and 52 MPa for 23Nb-3Fe, 26Nb-2Fe, 29Nb-1Fe, and 32Nb-0Fe, respectively. These values are relatively small and, more importantly, vary only within ~11 MPa across the alloy series. This variation is far below the experimental difference in yield strength, which decreases from 691 ± 13 MPa for 23Nb-3Fe to 468 ± 14 MPa for 32Nb-0Fe. Therefore, although grain-boundary strengthening may contribute to the overall strength level, it cannot explain the monotonic decrease in yield strength and hardness with decreasing Fe content. This is especially clear for 32Nb-0Fe, which has an intermediate β grain size and a Hall-Petch contribution comparable to that of 23Nb-3Fe, but exhibits the lowest yield strength. Thus, the mechanical trend can be more strongly controlled by alloy chemistry than by the β grain size.

The dominant compositional variable in the present alloy series is the progressive reduction in Fe content, from approximately 3 wt.% in 23Nb-3Fe to zero in 32Nb-0Fe. This is consistent with previous reports showing that Fe is a highly effective strength-enhancing solute in β-Ti alloys. Salvador et al. [46] investigated solute-lean Ti-Nb-Fe alloys, in which Nb was progressively reduced from 31 to 11 wt.% while Fe increased from 1.0 to 3.5 wt.%. Despite the reduction in Nb, the yield strength increased monotonically from 477 to 715 MPa, indicating that small additions of Fe can compensate for, and even exceed, the strengthening effect associated with much larger Nb contents [46]. A similar behavior was reported by Zhao et al. [47] for single-β Ti-Fe-Sn-Nb alloys, where Sn and Nb were fixed at 3 at.% and the Fe content was increased from 10 to 14 at.%. In that case, the compressive yield strength increased from approximately 1470 to 1880 MPa, and the authors associated this increase with higher β stability and a greater tendency for slip-dominated deformation rather than twinning, as rationalized using the Bo-Md map [47]. These studies support the interpretation that Fe affects not only the magnitude of solid-solution strengthening, but also the mechanical stability of the β matrix and the operative deformation mode.

Solid solution hardening (SSH) is commonly estimated using Eq. 4 as follows:

$$\sigma_{SS} = \left(\sum_i B_i^{3/2} X_i\right)^{2/3}, \qquad \text{Eq. 4}$$

where $B_i$ is the strengthening coefficient of a solute *i,* and its values were obtained from [43]. By using the experimental composition, the calculated $\sigma_{SS}$ of the 23Nb-3Fe, 26Nb-2Fe, 29Nb-1Fe, and 32Nb-0Fe alloys were 281, 214, 191, and 195 MPa, respectively. The difference between the values obtained for the contribution from SSH is still much lower than the experimental difference in yield strength. The role of Sn in SSH should be interpreted with caution because its strengthening effect depends strongly on the phase constitution and stability of the Ti matrix. Although Sn has been reported as an effective solid-solution strengthener in α-Ti alloys [48], this behavior cannot be directly transferred to metastable β-Ti alloys, where Sn may also suppress or delay ω formation, modify martensitic transformation tendencies, and increase the β lattice parameter. For example, Moraes et al. [49] reported that Sn additions to Ti-30Nb-xSn alloys reduced hardness, elastic modulus, compressive yield strength, and ultimate compressive strength up to intermediate Sn contents, while increasing ductility. They attributed this reduction to ω phase suppression/reduction and lattice parameter increase. In the present alloys, the β lattice parameter increased from approximately 3.261 Å in 23Nb-3Fe to 3.274 Å in 32Nb-0Fe. However, lattice expansion alone should not be taken as a direct measure of strengthening, because solid-solution strengthening depends on local size misfit, modulus misfit, chemical interactions, and solute-dislocation interactions. Indeed, the model used by Zhao et al. [43] assigns a higher intrinsic solid-solution coefficient to Sn (2303 MPa) than to Fe (1715 MPa), meaning that a purely coefficient-based calculation would not predict Fe-dominated strengthening. Therefore, we suggested that the coefficients proposed by [43] may overestimate the strengthening effect of Sn or underestimate that of Fe, as the decrease in strength from 23Nb-3Fe to 32Nb-0Fe is better explained by the systematic reduction in Fe content.

## 5. Summary and conclusions

This study demonstrates the significant benefits of using PMAPs, advanced PBF-LB processing with a remelting step, and subsequent heat-treatment to produce metastable β Ti-Nb-Fe-Sn alloys with precise compositional control and improved chemical and microstructural homogeneity. Based on experimental results and thermodynamic analysis, the main conclusions are:

1. Ti-Nb-Fe-Sn alloys were successfully produced by PBF-LB using PMAPs mixed with commercially pure Ti. After heat treatment at 1000 °C for 2 h followed by water quenching, all alloys exhibited low porosity, homogeneous SEM-BSE contrast, and a single β phase by XRD.
2. Thermodynamic calculations showed that solidification descriptors (Scheil freezing range, *Q*, and *P*) alone are insufficient to predict the final β grain-size evolution. The observed grain-size trend results from the combined effects of inherited solidification structure, transient phase transformations during heating, solute partitioning, and β grain-boundary mobility during the 1000 °C hold.
3. Hardness and yield strength decreased systematically with decreasing Fe and increasing Nb contents, ranging from 268 to 224 HV and from 691 to 468 MPa, respectively. The mechanical response was dominated by solid-solution strengthening, with additional contributions from grain-boundary strengthening.
4. The combination of master-alloy powders, layer remelting, and post-processing heat treatment provides an effective route for producing chemically homogeneous metastable β Ti-Nb-Fe-Sn alloys with tunable phase stability, β grain size, texture, and mechanical properties for biomedical applications.

## CRediT authorship contribution statement

**João Felipe Queiroz Rodrigues:** Data curation, Formal analysis, Investigation, Methodology, Visualization, Writing – original draft. **Kristína Bartha:** Data curation, Formal analysis, Investigation, Validation, Writing – original draft. **Mariano Casas-Luna:** Investigation, Writing – review and editing. **Gilberto Vicente Prandi:** Investigation, Writing – review and editing. **Márcio Sangali:** Investigation, Writing – review and editing. **Kateřina Ficková:** Investigation. **Jiří Kozlík:** Investigation, Writing

– review and editing. **Michaela Šlapáková:** Investigation, Writing – review and editing. **Martin Koller:** Formal analysis, Investigation, Writing – review and editing. **Adam Strnad:** Investigation, Writing – review and editing. **Josef Stráský:** Supervision, Validation, Writing – review and editing. **Miloš Janeček:** Conceptualization, Project administration, Resources, Writing – review and editing. **Rubens Caram:** Conceptualization, Project administration, Resources, Writing – review and editing.

## Acknowledgments

This work received funding from the Czech Science Foundation under the project 24-11074J. The Brazilian co-authors acknowledge the São Paulo State Research Foundation (FAPESP), grant #2023/13947-8, #2025/18810-6, #2018/18293-8, #2022/10049-6 for its financial support. Financial support by the Operational Programme Johannes Amos Comenius of the MEYS of the Czech Republic, within the frame of the project Ferroic multifunctionalities (FerrMion) (project No. CZ.02.01.01/00/22_008/0004591), co-funded by the European Union is also gratefully acknowledged.

## Data availability

The data that support the findings of this study are openly available in the Zenodo repository at https://doi.org/10.5281/zenodo.21993514 under the CC-BY 4.0 license.